\documentclass{PoS}
\title{Theory overview of charm physics}

\ShortTitle{Charm theory }

\author{\speaker{Svjetlana Fajfer }\thanks{Thanks to all collaborators. Work has been supported by ARRS, Slovenia }\\
       Department of Physics,
  University of Ljubljana, Jadranska 19
  and J. Stefan Institute, Jamova 39,  1000 Ljubljana, Slovenia\\
 \email{svjetlana.fajfer@ijs.si}}

\abstract{ Recent developments in charm physics within the standard model and beyond are reviewed. The precision study of leptonic charm meson decays accompanied with the constraints coming from  K, B and $\tau$ physics enables  to 
determine  constraints on  models of new physics. The standard model contributions in the $D^0 - \bar D^0$ system are still under investigation including operators of higher dimensions.
 New physics models such as the Littlest Higgs model or warped extra-dimensional models might  modify the amount of CP violation, although CP violating asymmetries are still predicted to be at most of the order $1 \%$. 
 The forward-backward asymmetry in the top anti-top pair production at the Tevatron is opening a new window in the study of new physics in the charm sector.
 A few of the models considered to explain the observed asymmetry might modify flavor changing neutral current mediated charm processes.
 New physics searches in  rare charm decays are discussed. It is pointed out that in most models new physics effects in rare charm decays are insignificant.}

\FullConference{The 13th International Conference on B-Physics at Hadron Machines - Beauty2011,\\
		April 04-08, 2011\\
		Amsterdam, The Netherlands}

\begin{document}
 
During the last few decades, flavor physics has continuously played an extremely important role in constraining models for new physics.  Contrary to  Flavor Changing Neutral Current (FCNC) processes of $K$ and $B$ mesons,  charm FCNC physics was not considered very attractive.  
%In the loop mediated $\Delta C =2$  and $\Delta C=1$ processes within Standard model,  the GIM mechanism plays very important role due to the down quarks in the loops and  the size of adequate CKM matrix elements. 
On the one hand, the charm quark sector provides us with an unique opportunity to test the predictions of the Standard Model (SM) and its extensions in the up quark sector.
On the other hand, the FCNC processes involving charm quarks are strongly affected by the presence of non-perturbative physics, since the masses of $s$ and $d$ quarks dominating the SM loop contributions are smaller than QCD scale. 
Additionally, in the $\Delta C =2$  and FCNC $\Delta C=1$ transitions GIM cancellations play a very important role. The down quarks are now intermediate states and when their masses are accompanied by appropriate CKM matrix elements often strong cancellations occur. 
This makes precise SM predictions and searches for New Physics (NP) rather difficult. 
Finally, in many theoretical studies the $b$ quark can be treated as static, while the $c$ quark is not heavy enough and corrections of higher orders in $1/m_c$ expansion become very important. 
 This lecture discusses recent results on charm meson leptonic and semileptonic decays, $D^0 - \bar D^0$ oscillations, CP violation in the charm system and rare charm decays.  Theoretical improvements within SM are considered and confronted with  recent experimental results. 
Concerning  search for NP, correlations between $\Delta C = 2$ and $\Delta C = 1$  processes are discussed, but also interplay of NP in $K$, $B$ and charm sectors.  Implications of NP searches involving charm for the recent intriguing results at the Tevatron in the top anti-top  pair production are discussed.  

\section{Leptonic  charm decays}

Charm meson leptonic decays  are very useful in providing an independent determination of the  CKM matrix elements $V_{cs}$ and $V_{cd}$, but also a possibility to test theoretical tools such as Lattice QCD and investigate the validity of heavy quark and operator product expansions in addressing perturbative and power corrections.  
Important nonperturbative parameters entering the discussion are the decay constants $f_{D_q}$ defined as $ < 0 | \bar q \gamma_\mu \gamma_5 c| D_q(p)> = i f_{D_q} p_\mu$.  
During the past six years  the experimental results from CLEO, Belle and BaBar had disagreed with Lattice QCD results (the so-called "$D_s$ puzzle"), stimulating many NP analyses. The results of the most recent lattice study of  Fermilab Lattice and Milk Collaboration~\cite{Lat-f} are however in good agreement with the latest HFAG world average experimental values~\cite{HFAG} 
\begin{eqnarray}
f_{D_s}^{\rm lat.}& =& 261 \pm 8 \pm 5 \enspace {\rm MeV}, \enspace  f_{D_s}^{\rm exp.} = 257.3\pm 5.3  {\rm MeV}, 
\nonumber\\
( f_{D_s} /f_{D})^{\rm lat.} &=& 1.19.\pm 0.01\pm 0.02 \enspace {\rm MeV}.
 \label{LatF}
 \end{eqnarray}
%Therefore  the lattice result is in a good agreement with experimental value. 
Although the disagreement between lattice and experimental  results has almost disappeared, leptonic charm decays still offer many useful constraints on NP. It was pointed out in~\cite{Kronfeld} that charged Higgs  or leptoquarks can contribute at the tree level. In scenarios with a charged Higgs interfering with the SM, useful bounds on $\tan \beta$ vs. the mass of the charged Higgs can be obtained, as presented in~\cite{akey,CKMfit}. %Akeyrod. 
%The same decay modes help to constrain the Two Higgs Doublet Dodel II giving the lower bound on the mass of charged Higgs, $m_{H^+} > 323$ GeV, while  any significant bound on $tan \beta$ was not found. 
The presence of weak triplet, doublet or singlet leptoquarks  in charm meson leptonic decays was investigated in~\cite{dorsner}.
The triplet leptoquarks with electric charge $-1/3$ can contribute to $D_q \to \ell \nu_\ell$ but also to $\tau \to \eta \mu$, 
$\tau \to \pi \mu$, $K \to \mu \nu_\mu$, $K^+ \to \pi^+\nu \bar \nu$, $K_L \to \mu^+ \mu^-$ and to the ratio $BR(\tau \to \pi \nu)/BR(\pi \to \mu \nu)$. Altogether, these observables exclude the triplet leptoquarks from affecting significantly the $D_q \to l \nu_l$ mode. The doublet leptoquark on the other hand, can mediate $K_L \to \mu^+ \mu^-$ and $\tau \to \eta \mu$. Then, the recent Belle upper bound on the branching ratio of $D^0 \to \mu^+ \mu^-$ excludes also the doublet leptoquarks from affecting $D_s\to \mu\nu$, while the tau mode is still unconstrained. The Lagrangian describing the singlet leptoquark  includes two terms. One of them can be matched to the R-parity violating minimal supersymmetric SM, where
the interaction term of a right-handed down squark to
quark and lepton doublets is present. However, both terms in the Lagrangian are necessary to to avoid bounds coming from $BR(K \to \mu \nu)/BR(\tau \to K \nu)$.  As in the doublet case, the lack of experimental information
on up-quark FCNCs involving tau leptons then
leaves open the verdict on the singlet leptoquark contribution
to the  $D_s \to \tau \nu$ decay width.

\section{Semileptonic  charm decays}

Inclusive charm meson decays can be used as a laboratory to test the operator product expansion techniques used in the extraction of $|V_{ub}|$  from inclusive semileptonic B decays. In this approach, a handle on the so-called weak annihilation contributions is very important to control theoretical uncertainties.  Recent analyses of perturbative $\alpha_s$ and $1/m_c$ power corrections 
in charm meson inclusive semileptonic decays show that there is no significant indication of weak annihilation contributions~\cite{jernej1,ligeti}.

Regarding the exclusive semileptonic modes, a recent lattice study of the semileptonic form factors in $D \to K \ell \nu_\ell$~\cite{HISQ}  finds for the vector form factor at zero lepton momentum squared $f_+(0) = 0.747\pm0.011\pm0.015$. Comparing this value to the recent experimental results,  the relevant CKM matrix element is found to be $|V_{cs}| = 0.961\pm0.011\pm 0.024$. 

\section{$D^0 - \bar D^0$ oscillations}

The $D^0-\bar D^0$ system is the only neutral meson system of valence up-type quarks.  Within the SM, the mixing of the two flavor eigenstates is driven by intermediate states of down-type quarks. Among them, the b quark contributions are negligible due to the smallness of the relevant CKM matrix elements. The masses of $s$ and $d$ quarks are however, smaller then the nonperturbative QCD scale $\Lambda_{QCD}$ implying that long distance contributions can dominate the dynamics of the $D^0 - \bar D^0$ system. 
A recent fit by HFAG~\cite{HFAG} for the four experimentally relevant observables yields
\begin{eqnarray}
x_D& = &\frac{\Delta M_D}{\Gamma_D} = (0.63^{+0.19}_{-0.20})\%, \enspace y_D = \frac{\Delta \Gamma_D}{\Gamma_D} = (0.75\pm0.12)\%, \nonumber\\
|\frac{q}{p}| &=&0.91^{+0.18}_{-0.16}, \enspace \Phi = (-10.2^{+9.4}_{-8.9})^o. 
\label{xD}
\end{eqnarray}
The  phase is defined as $\Phi =Arg(q/p) $.  $\Delta M$ and $\Delta \Gamma$ do not give any information on CP violation, while $|
q/p| \ne 1$ and $\Phi\ne \{ 0,\pm \pi \}$ would signal the presence of CP violation in the charm meson system.  However, this can be completely due to SM dynamics. 
There are presently two approaches to estimating the SM contributions: inclusive approach, based on the assumption of quark-hadron duality~\cite{Georgi,Ohl,Bigi}, and the exclusive approach, based on the study of individual decay channels contributing to 
the mixing of $D$ mesons~\cite{Donoghue,Falk,Cheng10}. Both approaches suffer from large theoretical uncertainties. 

The authors of~\cite{Lenz1} recently
investigated the SM CP violating effects in $D$-mixing within the inclusive approach, including  perturbative $\alpha_s$ as well as $1/m_c$ power corrections to the absorptive part of the mixing amplitude of neutral $D$ mesons.  In $D^0 - \bar D^0$ system $1/m_c$ corrections are suppressed by fourth power of $m_s$ in $x_D$ and by the sixth power in $y_D$.   It was shown that dimensions $6$ and $7$ operators lead to $10000$ smaller value of $\Gamma_{12}$ than the current experimental value.  Due to the strong GIM cancellation it might happen that the operators with the dimension $9$ or $12$ give dominant contributions. The authors  suggest that "CP violating effects in charm meson system of the order of $10^{-3}$ to $10^{-2}$ are an unambiguous sign of new physics"~\cite{Lenz1} . 

In the exclusive approach the contribution of the sum of intermediate  hadronic states is estimated. Recenly~\cite{Cheng10} the sum of all two body intermediate states was reinvestigated, based on the fact that $63\%$ of hadronic $D^0$ decays are two-body decays. The remaining multi-body hadronic decays contribute much less, particularly when cancellations among them are included into consideration.  The contributions considered in~\cite{Cheng10} are found to be primary  from the P P and V P modes leading to the estimates $x_D \sim 10^{-3}$ and  $y_D \sim 10^{-3}$. As pointed out already in~\cite{Petrov09}  $x_D \sim 1\%$ and $y_D \sim 1\%$ are possible within the SM. Due to the large uncertainty one cannot subtract these values from the experimental results and claim that the difference is due to new physics. Usually, the experimental values of $x_D$ and $y_D$ themselves are used as constraints on possible NP contributions, requiring that these do not exceed the experimental errors. 

If the masses of intermediate NP particles are above the electroweak scale one can integrate out these heavy degrees of freedom and consider effective $\Delta C= 2$ Lagrangian.  At the scale of the charm quark mass these contributions are modified by the strong interaction effects. The running QCD effects has been known for some time~\cite{Petrov07,Petrov08, Petrov07d}.  Taking operator mixing into account, the real and imaginary parts of the corresponding Wilson coefficients at the electroweak scale have been recently  constrained~\cite{Nir09,Buras10,Isidori10}.   

Among the possible new flavor structures beyond the SM, Minimal Flavor Violation (MFV) offers an interesting possibility that the NP shares the structure of the SM Yukawa terms. Promoting the Yukawa matrices to spurionic fields, transforming as $Y_u (3, \bar 3, 1)$ and $Y_d (3,  1, \bar 3)$ under the SM flavor group $G = SU(3)_Q \times SU(3)_U \times SU(3)_D$, the NP Wilson coefficients entering the effective Lagrangian, have to involve powers of $Y_d Y_d^\dag$ and $Y_u Y_u^\dag$. 
%General MFV \cite{zupan-gmfv} considers effective operators of the form  of the bilinears $\bar Q  Q$ with the insertions $\bar Q(Y_u Y_u^\dag)^n  Q$. 
Within a non-linear realization of MFV the authors~\cite{zupan-gmfv} found that the phase $\Phi$ 
in the $D^0 - \bar D^0$ system can be related to the CP-violating phase $\gamma$ entering B physics observables. 

Lately  a number of models of NP beyond MFV framework have been tested in the 
$D^0 - \bar D^0$ system.   In models where new flavor violating effects have been moved to the up sector, the leading constraints come from charm physics. 
% and $|M_{12}|\ne 0$, when the SM contribution  $(M_{12})_SM=  0$, or $(M_{12})_SM\in [-0.02, 0.02]$ ${\rm ps}^{-1}$. 
The fourth generation as the simplest SM extension in the flavor sector, has attracted a lot of attention  (c.f. ~\cite{Lenz2}).  In \cite{Lenz2}  the strongest bounds on $| V_{ub'} V_{cb'}|$ were derived from the $D$-mixing analysis
%from the $M_{12}$ 
for the mass of $b^\prime$ ranging from $200 - 500$ GeV. 
%Correlations between constraints coming from low energy phenomenology of  $B$ and $K$ mesons, and charm meson physics, used in almost all of the above-mentioned studies, have recently been questioned in~\cite{??}. 

\section{Implications for the top quark sector}

The measured forward-backward asymmetry in the $t \bar t$ production and especially at high invariant  masses of the $t \bar t$  system at the Tevatron 
 offer new opportunities for new physics searches at low energies.  It is very intriguing that only the  measured asymmetries  deviate from  SM  predictions, while the measured production cross section and invariant mass spectrum are in reasonably good agreement with SM predictions.  Among many models (see e.g.~\cite{Gresham}),  the  exchange of a colored weak singlet scalar in the u-channel and its interference with the SM  contribution can accommodate all relevant observables measured at the Tevatron.  Such state appears in some of the grand unified theories and its interactions with the up-quarks are purely antisymmetric in flavor space. The resulting impact on charm and top quark physics has been systematically investigated in~\cite{our-LQt}. It turns out that one of the most important constraints on the relevant couplings comes from the experimentally measured observables related to $D^0 - \bar D^0$ oscillations
 %, as well as di-jet and single top production measurements at the Tevatron.  
 While the $t\bar t$ phenomenology requires the $u- t$ coupling with the exotic scalar state of the order 1, the corresponding $c-t$ coupling entering the box mediated 
 $c \bar u \leftrightarrow \bar u c$,  is then constrained by $D$-mixing to be two orders of magnitude smaller. In a class of grand unified models it has been demonstrated that these constraints affect the up-quark Yukawa couplings leading to a lopsided up-quark mass matrix~\cite{our-LQ-up} . 
 Models with a color sextet weak singlet scalar can also explain observed asymmetries at the Tevatron without significant modifications of the cross section for the $t \bar t$ production (see e.g. \cite{Gresham}). However, in $D^0 - \bar D^0$ oscillations the color sextet can contribute already at the tree level~\cite{chen10}, resulting in even more severe constraints on the $u - c$ sextet coupling.  Likewise, a new  flavor changing $Z^\prime$  gauge boson is also successful in accomodating the Tevatron $ t \bar t$ observables. However, the $Z^\prime$ mediation of the  $D^0 - \bar D^0$  oscillations at the tree level again requires the $cu Z^\prime$ coupling to be orders of magnitude smaller than the $t  u Z^\prime$. On the theoretical side such alignment imposes questions about the underlying flavor structure of such models.
%Generally if the need for NP in $t \bar t$ asymmetry observed at Tevatron will persist one should understand what are the implications in the down-like sector and also are there some models NP which contain new particles  responsible for the anomalous behavior in $t \bar t$ production. The GUT model with color triplet weak singlet scalar described in \cite{our-LQt,our-LQ-up} is one of such candidates. 
%The modification of the Randall - Sundrum model  with heavier Kaluza-Klein excitation with the mass $\sim 8.5$ TeV might lead to satisfactory explanation of the Tevatron excess in $t \bar t$ production \cite{Del,Dju} also.
 
 \section{New CP violation in charm}

The Littlest Higgs model with T-parity (LHT) offers an appealing  solution to the Higgs hierarchy problem  without causing problems with the electroweak constraints. It introduces a discrete symmetry called T-parity, under which the new particles can be odd so that they contribute to processes of SM particles only 
 at the loop level. The model contains three new families of mirror quarks and consequently two additional CP violating phases. 
 %In~\cite{BigBuras} it has been assumed that $\Gamma_{12}$ is dominated by SM, while $M_{12}$ can receive large non SM  contributions. 
Parameters describing these new degrees of freedom can then be related to CP asymmetries in $K$, $B$ and $D$ decays of \cite{BigBuras} found that LHT model can produce observable CP violating effects in the charm system. In the case of nonleptonic neutral charm meson decay to $K_S \Phi$ two CP violating quantities $S_f$ and $a_{SL} (D^0)$ offer tests of LHT dynamics~\cite{Bigi-CP}.  It is interesting that indirect CP violation in the nonleptonic decays $D^0 \to \pi^+ \pi^-$, $D^0 \to K^+ K^-$ and $D^0 \to \Phi K_S$ will create the same asymmetries. If experiment will find different CP asymmetries, then they have to be the result of direct CP violation. The authors of~\cite{Bigi-CP} noticed that  an enhancement factor of two in $a_{CP}^{dir}$ is possible for the unconstrained set of parameters. 

On the other hand, in the Randall - Sundrum model studied in~ \cite{Bauer10} the CP violating parameters can reach $S_{\Phi K_S}\in [0.25,0.15]$ and $a_{SL} (D^0)\in[-0.75, 0.4]$ for the set of model parameters constrained by $|\epsilon_K|$, $Z^0 \to \bar b b$, and $B_d - \bar B_d$ oscillations and assuming $\Gamma_{12}^D= -0.02$ and $ 0.02$ ${\rm ps}^{-1}$.

\section{Rare charm decays}

In the SM the contribution coming from penguin diagrams in 
$\rm c\to u\gamma$ transition is
strongly GIM suppressed giving a 
branching ratio of order $10^{-18}$~\cite{burdman1}, while the 
QCD corrected
effective Lagrangian gives $\rm BR(c\to u\gamma)\simeq3\times10^{-8}$
\cite{Greub}. 
A variety of models beyond the SM have been
investigated and it was found that the gluino exchange diagrams~\cite{Sasa} within the general minimal supersymmetric SM (MSSM) might lead to significant 
enhancements. With the updated constraints it was found that  ~\cite{Nejc} 
$\rm{\Gamma (c\to u\gamma)}/{\Gamma_{D^0} }\le 8 \times 10^{-7} $.

Likewise, within the SM the $c\to u l^+l^-$ amplitude is given by the $\gamma$ and $Z$ penguins
and $W$ box diagrams at one-loop electroweak level. It is dominated by the light quark contributions in
the loops.  
The leading order rate for the inclusive $c\to u l^+l^-$ calculated within 
 SM~\cite{Singer} was found to be suppressed by QCD corrections~\cite{burdman2}. 
Inclusion of renormalization group equations  for  Wilson coefficients 
gave an additional significant 
suppression~\cite{Jure} leading to the rates  
$\Gamma(c\to ue^+e^-)/\Gamma_{D^0}=2.4\times 10^{-10}$ and
$\Gamma(c\to u\mu^+\mu^-)/\Gamma_{D^0}=0.5\times 10^{-10}$.   
These transitions are largely driven by virtual photon contributions at low dilepton mass.  
The leading 
contribution to $c\to ul^+l^-$ in the MSSM with the conserved R parity 
comes from one-loop diagrams~\cite{burdman2,Sasa} with 
gluino and squarks in the loop. 
It proceeds via virtual photon  
and significantly enhances the $c\to ul^+l^-$ 
spectrum at small dilepton mass $m_{ll}$. 
The   effects of the extra heavy up vector-like quark models on the decay spectrum of $D^+ \to \pi^+ l^+ l^-$ and $D^+_s  \to K^+ l^+ l^-$ 
 decays were also considered~\cite{Nejc}. It was found that there is a tiny increase of the differential decay rate in the region of large dilepton mass. %The R-parity violating supersymmetric model can also modify short distance dynamics in  $c\to u l^+l^-$decays.The relevant parameters were constrained using current upper bound on the $D^+ \to \pi^+ l^+ l^-$  decay rate. Present bounds still allow small modification of the standard model differential decay rate distribution.

Recently $D^0 \to \gamma \gamma$ and $D^0 \to l^+  l^-$  were reconsidered in~\cite{Bigi-Stef}.  The result for the short and long distance contributions are 
%\begin{eqnarray}
$BR_{SD}^{2-loops} (D^0 \to \gamma \gamma) \simeq (3.6-8.1) \times 10^{-12}$ , %\nonumber\\
$BR_{SD}^{2-loops} (D^0 \to \gamma \gamma) \simeq (3.6-8.1) \times 10^{-12}$.
%\label{r}
%\end{eqnarray}
Current bound is quite weak with $BR_{exp}(D^0 \to \gamma \gamma) \sim 2.7  \times 10^{-5}$ and it seems that real improvements can be achieved only at the Super Flavor or the Super Tau-Charm factory. 
Short distance contributions in $D^0 \to l^+  l^-$ decay lead to a very suppressed branching ratio in the SM. Therefore it is natural to consider it is as an ideal testing ground for NP effects.  Reference ~\cite{Bigi-Stef}  considered contributions coming from $ \gamma \gamma$ intermediate states due to long distance dynamics in $D^0 \to \gamma \gamma$  arriving at the value $BR(D^0 \to \mu^+  \mu^-) \sim (2.7 - 8) \times 10^{-13}$. According to their calculations, the LHT model can enhance the branching ratio for a factor of 2. 

In~\cite{Petrov07,Petrov09} an operator product expansion has been used to obtain the effective Hamiltonian and explore the correlations of the NP in $D^0 - \bar D^0$ oscillations and $D^0 \to \mu^+ \mu^-$. 
Although NP contributions are dominant in this decay channel they in the best case (vector-like quark models) \cite{Petrov07,Petrov09} increase the rate by a factor of $100$. 
The experimental bounds on $D^0 \to \ell^+  \ell^-$ have recently been  improved by the Belle collaboration to $BR(  D^0 \to \mu^+  \mu^-)\le 1.4 \times 10^{-7}$ and $BR(  D^0 \to e^+  e^-)\le 7.9  \times 10^{-8}$~\cite{Staric}.

\section{Conclusions} 
The accumulated data on charm physics give new insights into SM contributions and further improvements of the theoretical tools are still possible. In particular the complete knowledge of the long distance contributions in  the $x_D$ and $y_D$  observables in $D^0 - \bar D^0$ oscillation is still lacking. Therefore, NP constraints could still be improved. 
The search for CP violating signals has progressed and a lot of models of new physics have predicted increase of CP violating asymmetries relative to SM results. However, they all of the order $1 \%$. 
The rare decay modes can be enhanced by at most two orders of magnitude in most SM extensions, still far bellow the current experimental precision.
The anomalous behavior of the $ t  \bar t$ production at Tevatron opened new paths in the study of NP effects in the charm sector. If the measured asymmetries remain, this will require new insight into NP in charm as well. With current searches at LHCb and with future planned facilities including the charm physics program,  we expect a stimulating and fruitful era for charm. 

%\begin{eqnarray} 

\end{document}